\documentclass{ws-ijmpd}
\usepackage[super,compress]{cite}
\usepackage{url}
\begin{document}

\markboth{Authors' Names}
{Instructions for Typing Manuscripts (Paper's Title)}

%
\catchline{}{}{}{}{}
%

\title{SUPERNOVA CONSTRAINTS ON HIGHER-DIMENSIONAL COSMOLOGY WITH A PHANTOM FIELD}

\author{JAMES OVERDUIN\footnote{Corresponding author (joverduin@towson.edu). Also at Department of Physics and Astronomy, Johns Hopkins University, 3400 North Charles Street, Baltimore, Maryland, 21218, U.S.A.}}

\address{Department of Physics, Astronomy and Geosciences, 8000 York Road, Towson University, Towson, Maryland, 21252, U.S.A.\\
}

\author{NATHAN PRINS}

\address{Department of Physics, Astronomy and Geosciences, 8000 York Road, Towson University, Towson, Maryland, 21252, U.S.A.\\
}

\author{JOOHAN LEE\footnote{Also at Department of Physics, University of Seoul, Seoul 130-743, Korea.}}

\address{Department of Physics, Astronomy and Geosciences, 8000 York Road, Towson University, Towson, Maryland, 21252, U.S.A.\\
}

\maketitle


\begin{abstract}
We use observational data on the magnitude-redshift relation for Type~Ia supernovae together with constraints on the ages of the oldest stars to rule out a higher-dimensional extension of General Relativity with a negative kinetic-energy scalar field.
This theory is of considerable physical interest because it produces accelerated expansion at both early and late times with a single new field, as in quintessential inflation scenarios.
It is also of mathematical interest because it is characterized by an analytic expression for the macroscopic scale factor $a(t)$.
We show that cosmological solutions of this theory can be usefully parametrized by a single quantity, the lookback time $\tau_{\text{tr}}$ corresponding to the transition from deceleration to acceleration.
Supernovae data from the recently released Supernova Cosmology Project Union~2.1 compilation single out a narrow range of values for $\tau_{\text{tr}}$.
In the context of the theory, however, these same values of $\tau_{\text{tr}}$ imply that the universe is much older than the oldest observed stars.
\end{abstract}

\keywords{dark energy; supernovae; scalar fields; extra dimensions.}

\ccode{04.50.Cd,95.36.+x,97.60.Bw,98.80.-k}

\section{Introduction}

The Universe is widely believed to have undergone a period of accelerated expansion at early times (driven by a scalar field known as the inflaton), followed by a period of decelerating expansion (driven by matter and radiation), and then by a transition to renewed acceleration that will likely continue forever (driven by dark energy).
Dark energy is often identified with Einstein's cosmological constant, but there are longstanding theoretical objections to this quantity \cite{Weinberg1989,KraghOverduin2014}.

A possibly more palatable alternative is to associate dark energy with a dynamical or ``variable cosmological constant'' \cite{OverduinCooperstock1998}, often in the form of another scalar field known as quintessence \cite{CaldwellEtAl1998}.
But this introduces two new scalar fields into cosmology, when only one quite different scalar (the Higgs boson) has actually been observed in nature.
Motivated by Occam's razor, there has been a recent push to unify the inflaton and dark energy in a single theory.
Models of this type go by the generic name quintessential inflation \cite{PeeblesVilenkin1999,LeeEtAl2014}.

Here we consider one recent quintessential inflation-type theory \cite{HongEtAl2008} based on a single scalar field in a ten-dimensional extension of General Relativity with a cosmological constant, as motivated by superstring and other approaches to unification \cite{OverduinWesson1997}.
This theory gives rise to accelerated expansion at both early and late times, with an intervening epoch of deceleration, if the scalar field is of the phantom type.
Phantom fields are characterized by negative kinetic energy and an equation of state with $p<-\rho c^2$ \cite{Caldwell2002}.
Despite these unusual properties, they are consistent with, and even arguably preferred by observational constraints on the equation-of-state parameter $w=p/\rho c^2$ \cite{DunkleyEtAl2009}.

We emphasize that our intent is not to promote or refine this particular theory; it is to confront it with observational data.
The same method developed here could also be applied to other theories in which the cosmological scale factor is known either analytically or numerically as a function of cosmic time \cite{YoshidaShiraishi1991,OverduinEtAl2007}.

\section{Higher-dimensional phantom cosmology}

\begin{figure*}[t!]
\centering
\includegraphics[width=\textwidth]{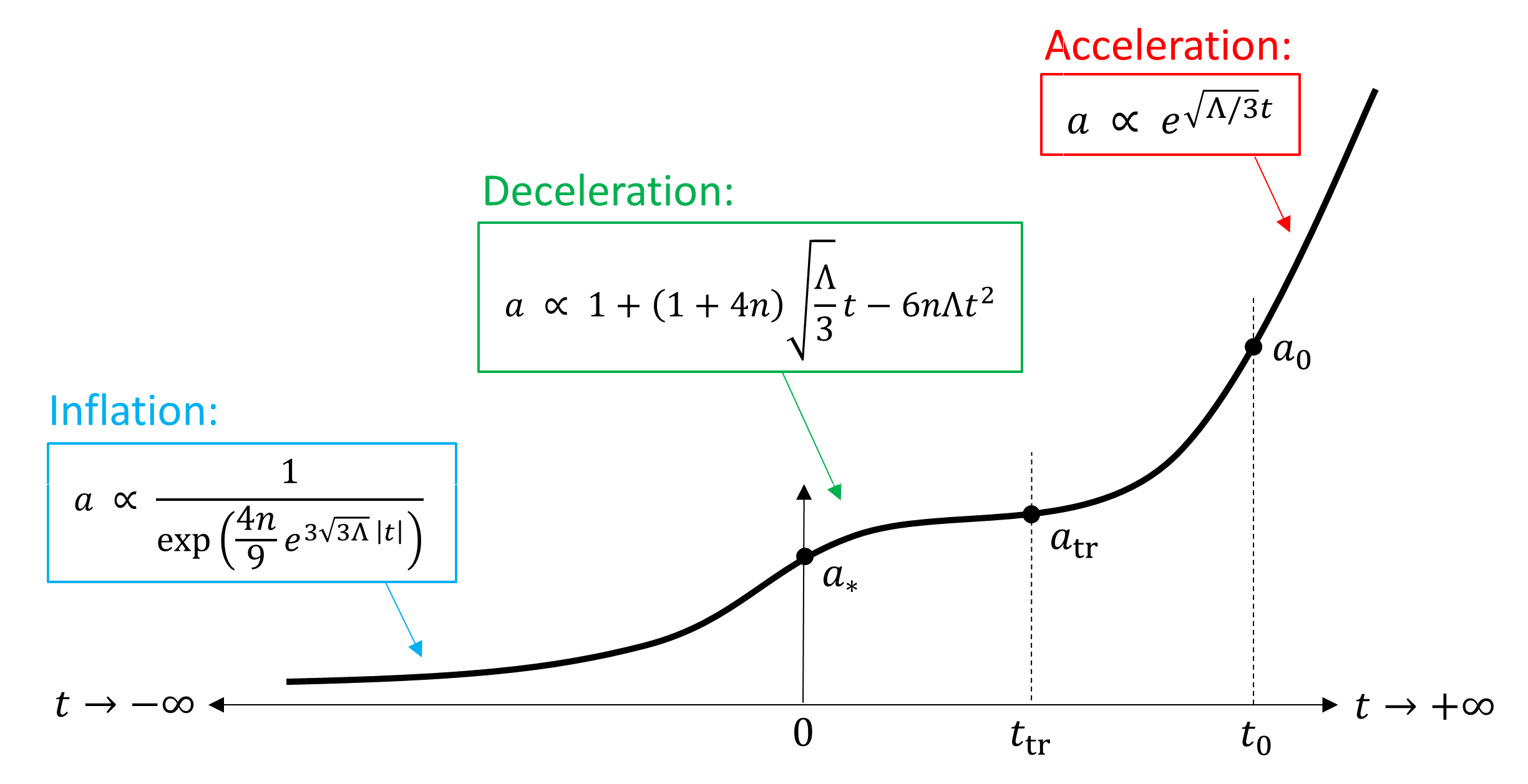}
\caption{Qualitative behavior of the macroscopic scale factor $a$ as a function of time in the model of Hong et al. \cite{HongEtAl2008}, showing inflation at early times, acceleration at late times, and deceleration at intermediate times.}
\label{fig-scaleFacQual}
\end{figure*}

The action of the theory reads\cite{HongEtAl2008}
\begin{equation}
S = \int d^{10} x \sqrt{-g} \Bigl( -\frac{1}{2} R + \frac{1}{2} g^{MN} \partial_M \sigma \partial_N \sigma - V(\sigma) - \bar{\Lambda} \Bigr) \; ,
\label{Action}
\end{equation}
where $\sigma$ is the scalar field, $V$ its potential, $\bar{\Lambda}$ the 10D cosmological constant, and upper-case Latin indices $M,N$ range over 0-9.
Variation leads to the equations of motion
\begin{eqnarray}
& & R_{MN} - \frac{1}{2} g_{MN} R - g_{MN} \bar{\Lambda} = -T_{MN} \; , \nonumber \\
& & \frac{1}{\sqrt{-g}} \partial_M \left( \sqrt{-g} \, g^{MN} \partial_N \sigma \right) + \frac{\partial V}{\partial \sigma} = 0 \; .
\label{eqnofmotion}
\end{eqnarray}
For simplicity one assumes that both the macroscopic 4D spacetime and the compact 6D space have Friedmann-Robertson-Walker (FRW) form, with scale factors $a(t)$ and $b(t)$, so that
\begin{equation}
g_{MN} = 
\begin{pmatrix}
-1 & 0 & 0 \\
0 & a^{2}(t) & 0 \\
0 & 0 & b^{2}(t) \\
\end{pmatrix} \; .
\label{metricform}
\end{equation}
Solving Eqs.~(\ref{eqnofmotion}) for the vacuum case ($T_{MN}=0$) with a vanishing potential $(V(\sigma)=0)$, Hong et al. \cite{HongEtAl2008} were able to find a class of new exact solutions for $a(t)$, $b(t)$ and $\sigma(t)$.
The macroscopic scale factor $a(t)$ undergoes de~Sitter-like exponential expansion at late times and doubly-exponential inflation at early times (driven by the phantom field), with a decelerating phase at intermediate times (under the influence of a matter-like term associated with the compact extra dimensions):
\begin{equation}
a(t)=a_{\ast} \exp \left[ \frac{1}{6} \sqrt{\bar{\Lambda}}\,t - \frac{4n}{9} \left( e^{-\frac{3}{2} \sqrt{\bar{\Lambda}}\,t} -1 \right) \right] \; .
\label{JoohanScaleFactor}
\end{equation}
Here the constant $n=(\dot{a}/a-\dot{b}/b)_{\ast}/\sqrt{\bar{\Lambda}}$ is effectively a free parameter of the theory, and $a_{\ast}$ is the scale factor at time $t=0$.
In contrast to Ref.~\refcite{HongEtAl2008}, we denote quantities at this time with a subscript ``$\ast$'' rather than ``0'' in order to avoid confusion with standard notation in big-bang cosmology, where the subscript ``0'' is reserved for quantities measured at the present time.
We follow the standard convention here, and note that the time $t=0$ has no physical significance in the model of Hong et al. \cite{HongEtAl2008}.
In particular, it does not mark the time when $a\rightarrow 0$, as in standard FRW models.
Rather, $a$ approaches arbitrarily close to zero as $t\rightarrow -\infty$ in the past direction.
For realistic choices of parameters this must of course be observationally indistinguishable from the standard picture, as we require below.
The behavior of $a(t)$ based on Eq.~(\ref{JoohanScaleFactor}) is shown schematically in Fig.~\ref{fig-scaleFacQual}, where $\Lambda=12\bar{\Lambda}$ is the standard (4D) cosmological constant.

\begin{figure*}[t!]
\centering
\includegraphics[width=\textwidth]{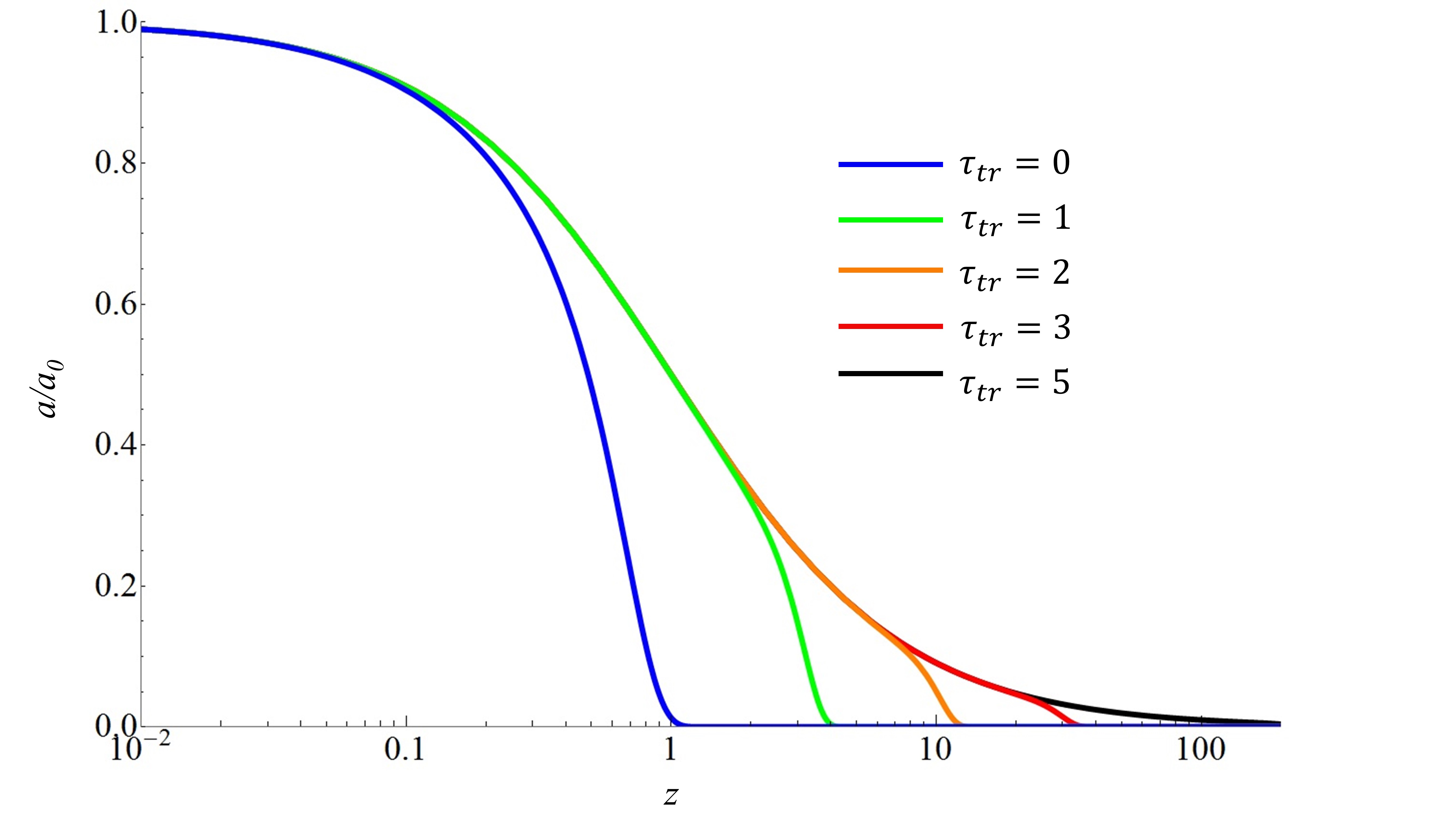}
\caption{Numerical plots showing the quantitative evolution of $a/a_0$ as a function of redshift for various values of $\tau_{\text{tr}}$, the lookback time corresponding to the most recent transition from deceleration to acceleration.}
\label{fig-scaleFacQuant}
\end{figure*}

Equation~(\ref{JoohanScaleFactor}) contains three constants whose significance is largely theoretical: $a_{\ast}$, $\sqrt{\bar{\Lambda}}$ and $n$.
To make better contact with observation, we first re-express the scale factor relative to its value $a_0=a(t_0)$ as measured, say, from the time when the scale factor was small enough, and the corresponding photon temperature hot enough, for standard big-bang nucleosynthesis to occur.
Eq.~(\ref{JoohanScaleFactor}) then becomes
\begin{equation}
\frac{a(t)}{a_0} = \exp\Bigl[ -\frac{1}{6}\sqrt{\bar{\Lambda}}(t_0-t) -\frac{4n}{9}e^{-\frac{3}{2}\!\sqrt{\bar{\Lambda}}t_0}\left(e^{+\frac{3}{2}\!\sqrt{\bar{\Lambda}}(t_0-t)}-1\right)\Bigr] \; .
\label{scaleFactor}
\end{equation}
Differentiating with respect to $t$, we find that
\begin{equation}
H(t)\equiv\frac{\dot{a}(t)}{a(t)}=\left(\frac{1}{6}+\frac{2n}{3}e^{-\frac{3}{2}\!\sqrt{\bar{\Lambda}}t}\right)\!\!\sqrt{\bar{\Lambda}} \; .
\label{hubbleParameter}
\end{equation}
The asymptotic limit of this equation allows us to replace $\sqrt{\bar{\Lambda}}$ with
\begin{equation}
\sqrt{\bar{\Lambda}}=6H_{\infty} \; .
\label{lambdaDefn}
\end{equation}
Differentiating $a(t)$ twice with respect to $t$ and equating to zero to find the inflection points, we discover that
\begin{equation}
4ne^{\frac{3}{2}\!\sqrt{\Lambda}t}=4ne^{-9H_{\infty}t}=\frac{1}{2}(7\pm3\sqrt{5}) \; .
\end{equation}
Hence we can replace $n$ by
\begin{equation}
n=\frac{\alpha}{4}e^{9 H_{\infty} t_{\text{tr}}} \;\;\; \mbox{ where } \;\;\;
\alpha\equiv\frac{1}{2}(7-3\sqrt{5}) \; ,
\label{nDefn}
\end{equation}
corresponding physically to the most recent inflection point; i.e., the time $t_{\text{tr}}$ of transition from deceleration to acceleration.
Putting Eqs.~(\ref{lambdaDefn}) and (\ref{nDefn}) into Eq.~(\ref{scaleFactor}), we find that the scale factor takes the form
\begin{equation}
\frac{a(\tau)}{a_0} = \exp \left[ -h_{\infty}\tau - \frac{\alpha}{9} e^{-9h_{\infty}\tau_{\text{tr}}} \left(e^{9h_{\infty}\tau}-1\right) \right] \; ,
\label{scaleFactorFinal}
\end{equation}
where we have defined $h_{\infty}\equiv H_{\infty}/H_0$ and replaced time $t$ with lookback time in Hubble units via
\begin{equation}
\tau\equiv H_0(t_0-t) \; .
\end{equation}
Expressed in this form, the scale factor is characterized by the known constant $\alpha$ plus two parameters $h_{\infty}$ and $\tau_{\text{tr}}$.
The asymptotic Hubble expansion rate $h_{\infty}$ is determined by Eqs.~(\ref{hubbleParameter}) and (\ref{nDefn}), which combine to give
\begin{equation}
h_{\infty} = \left( 1 + \alpha \mathrm{e}^{-9 h_{\infty} \tau_{\text{tr}}} \right)^{-1} \; .
\label{hfinal}
\end{equation}
This is a transcendental equation that may be solved numerically for $h_{\infty}$ in terms of $\alpha$ and $\tau_{\text{tr}}$.
Alternatively, we find that Eq.~(\ref{hfinal}) is well fit by
\begin{equation}
h_{\infty}=1-\beta \exp(-\gamma\tau_{\text{tr}}) \; ,
\label{hinftyApprox}
\end{equation}
with $\beta=0.127$ and $\gamma=8.377$ over the range of transition times considered here \cite{PrinsEtAl2014}.
This leaves us with only one adjustable parameter in the theory: the transition lookback time $\tau_{\text{tr}}$ (in Hubble units).
The evolution of $a/a_0$ is plotted as a function of redshift for a wide range of values of $\tau_{\text{tr}}$ in Fig.~\ref{fig-scaleFacQuant}.

\section{Magnitude-redshift test}

To test the theory, we use the magnitude-redshift relation for Type~Ia supernovae (SNeIa) \cite{GoobarLeibundgut2011} together with observational data for 580 SNeIa in the SCP (Supernova Cosmology Project) Union~2.1 compilation \cite{SuzukiEtAl2012}.
Two kinds of quantities are plotted as functions of redshift and compared with this data: distance modulus $\mu(z)$ and the magnitude residual $\Delta m(z) = m(z)-m_{\text{fid}}(z)$, where ``fid'' refers to a fiducial or reference cosmology, here standard $\Lambda$CDM with WMAP 9-year values for $h_0=0.70$, $\Omega_{\Lambda}=0.72$ and $\Omega_M=1-\Omega_{\Lambda}$ \cite{HinshawEtAl2013}.

Distance modulus $\mu(z)$ and apparent magnitude $m(z)$ are related to the absolute magnitude $M$ and the luminosity distance $d_L$ of the source by
\begin{equation}
\mu(z) = m(z) - M = 40 + 5 \log_{10} \left[ \frac{d_{L}(z)}{\mbox{Gpc}} \right] \; ,
\label{distmod}
\end{equation}
where $M$ is equal to apparent magnitude at a distance of 10~pc .
Luminosity distance is
\begin{equation}
d_L(z) = \frac{c (1+z)}{H_0\,\sigma_k} S_k\left[ \sigma_k \int_{0}^{z} \frac{H_0}{H(z)} \,dz \right] \; ,
\label{LumDist}
\end{equation}
Here the constant $\sigma_k$ and function ${\cal S}_k$ are defined so that
$\sigma_k\equiv\{\sqrt{\Omega_M+\Omega_{\Lambda}-1},1,\sqrt{1-\Omega_M-\Omega_{\Lambda}}$ and
${\cal S}_k[X]\equiv\{\sin X,X,\sinh X\}$ respectively for FRW models with $k=\{+1,0,-1\}$.
For these standard models, we use the Friedmann-Lema\^{i}tre equation for the Hubble expansion rate
\begin{equation}
\frac{H_{\text{FRW}}(z)}{H_0} =\sqrt{\Omega_M(1+z)^3+\Omega_{\Lambda}-\Omega_C(1+z)^2} \; ,
\label{Hfrw}
\end{equation}
where $\Omega_C\equiv\Omega_M+\Omega_{\Lambda}-1$.

\begin{figure}[t!]
   \centering
   \includegraphics[width=\columnwidth]{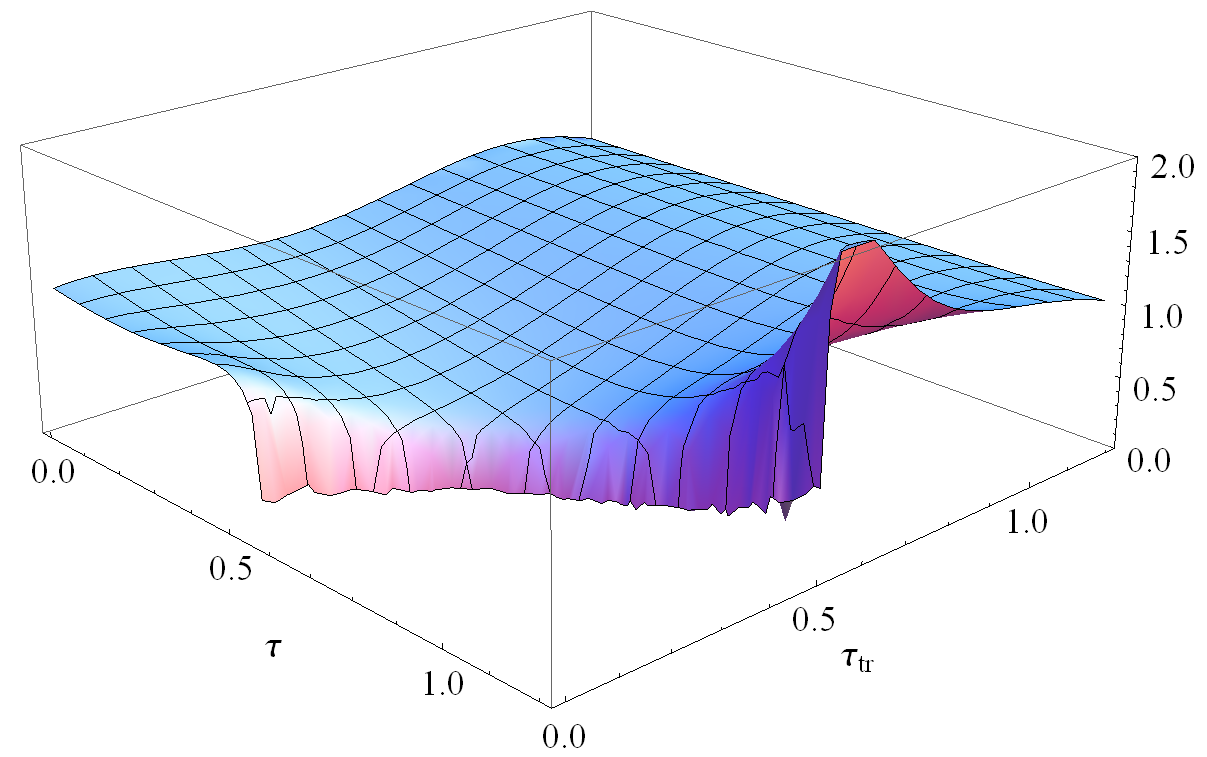}
   \caption{Ratio of $\ln(1+z)$ as defined by the approximation Eq.~(\ref{ln(1+z)fit}) relative to the definition in Eq.~(\ref{ln(1+z)}), plotted over the parameter space defined by physically plausible transition times $\tau_{\text{tr}}$ and the lookback times $\tau$ spanned by our supernova sample.}
\label{fig-numFit}%
\end{figure}

There are no ready analogs for $\Omega_M$ or $\Omega_{\Lambda}$ in the model of Hong et al. \cite{HongEtAl2008}, so the Hubble expansion rate cannot be calculated from Eq.~(\ref{Hfrw}) in that theory.
Instead, we change variables according to
\begin{equation}
\frac{dz}{H(z)} = -\frac{dt}{\left[a(t)/a_0\right]} \; ,
\end{equation}
which follows from the definitions of redshift, $z=[a_0-a(t)]/a(t)$, and the Hubble parameter $H\equiv\dot{a}/a$.
Expressed in terms of lookback time, we then find that Eq.~(\ref{LumDist}) takes the following form for spatially flat models ($k=0$):
\begin{equation}
d_L(z) =\frac{c (1+z)}{H_0} \int_{\tau_0}^{\tau(z)} \left[\frac{a(\tau^{\prime})}{a_0}\right]^{-1} \,d\tau^{\prime} \; ,
\end{equation}
where $\tau(z)$ is the lookback time corresponding to redshift $z$.
Inserting Eq.~(\ref{scaleFactorFinal}) into this integral, we obtain
\begin{equation}
d_L(z)=\frac{c (1+z)}{H_0} \int_0^{\tau(z)} \mathrm{exp}[h_{\infty} \tau^{\prime} + \frac{\alpha}{9} \mathrm{e}^{-9 h_{\infty} \tau_{\text{tr}}} (\mathrm{e}^{9 h_{\infty} \tau^{\prime}} - 1)] \,d\tau^{\prime} \; .
\label{LumDistsigma}
\end{equation}
This expression can then be used in Eq.~(\ref{distmod}) to calculate the distance modulus, which in turn allows us to find the magnitude residuals. 
To find an expression for the limit of integration $\tau(z)$ we note from Eq.~(\ref{scaleFactorFinal}) that
\begin{equation}
\ln(1+z) = \ln\left[\frac{a_0}{a(\tau)}\right] = h_{\infty} \tau + \frac{\alpha}{9} \mathrm{e}^{-9 h_{\infty} \tau_{\text{tr}}} (\mathrm{e}^{9 h_{\infty} \tau} - 1) \; .
\label{ln(1+z)}
\end{equation}
After some experimentation, we find that the inverse of this relation is well described by a relation of the form
\begin{equation}
\tau(z) = \tau_{\infty}(\tau_{\text{tr}})\left[ 1 - \frac{1}{(1+z)^{\,\kappa(\tau_{\text{tr}})}}\right] \; ,
\label{tauzfit}
\end{equation}
where
\begin{equation}
\tau_{\infty}(\tau_{\text{tr}}) = c_0+c_1\exp(c_2\tau_{\text{tr}}) \;\;\; \mbox{ and } \;\;\; \kappa(\tau_{\text{tr}}) = c_3\exp(-c_4\tau_{\text{tr}}^2) \; .
\end{equation}
A good fit is obtained over the redshifts and transition times considered here with $c_0=0.527$, $c_1=0.0240$, $c_2=6.32$, $c_3=1.87$ and $c_4=3.09$.
(By comparison, a first-order Taylor approximation to Eq.~(\ref{ln(1+z)}) assuming $z\ll1$, $h_{\infty}(\tau-\tau_{\text{tr}})\ll1$ and $h_{\infty}\tau_{\text{tr}}\ll1$, so that $\tau(z)\approx z/(1+\alpha)h_{\infty}$, is not sufficient; in this approximation the theory reduces to pure de~Sitter inflation \cite{PrinsEtAl2014}.)
Eq.~(\ref{tauzfit}) is both simple and physically plausible, in that it shows the standard cosmological power-law dependence on $(1+z)$.
To check the fit, we invert Eq.~(\ref{tauzfit}):
\begin{equation}
\ln(1+z)=-\frac{1}{\kappa(\tau_{\text{tr}})}\ln\left[1-\frac{\tau}{\tau_{\infty}(\tau_{\text{tr}})}\right] \; ,
\label{ln(1+z)fit}
\end{equation}
and plot the ratio of Eqs.~(\ref{ln(1+z)fit}) and (\ref{ln(1+z)}) in Fig.~\ref{fig-numFit}.
Figure~\ref{fig-numFit} confirms that the approximation is excellent over all  relevant values of $\tau$ and $\tau_{\text{tr}}$, failing only in the unphysical corner of the parameter space where lookback times are large but the transition to acceleration occurred recently (i.e., to cases where one is attempting to ``look past the big bang'').

\section{Results and discussion}

\begin{figure*}[t!]
   \centering
   \includegraphics[width=\textwidth]{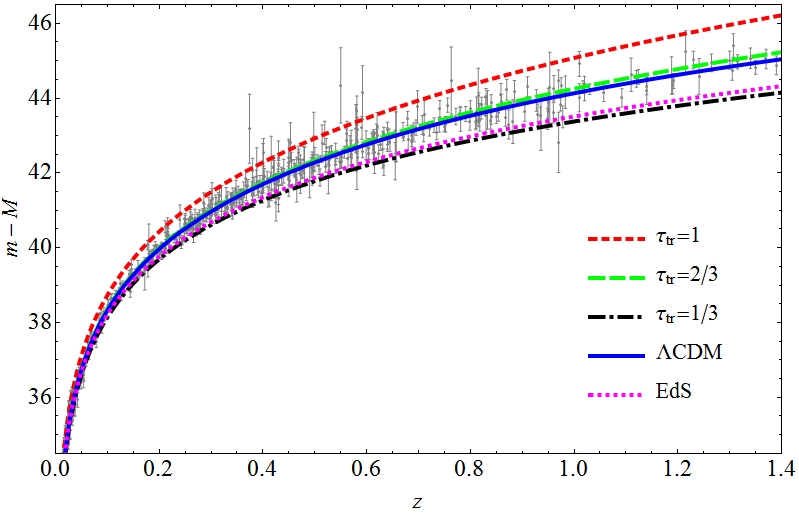}
   \caption{Distance modulus versus redshift for 580 SNeIa (SCP Union~2.1 compilation; grey points), plotted together with theoretical predictions for the model of Hong et al. \cite{HongEtAl2008} with three representative values of transition time $\tau_{\text{tr}}$ (dashed and dash-dotted lines) as well as the fiducial $\Lambda$CDM model (solid blue) and the Einstein-de~Sitter model with $\Omega_M=1,\Omega_{\Lambda}=0$ (dotted magenta) for reference.}
\label{fig-distMod}%
\end{figure*}

\begin{figure*}[t!]
   \centering
   \includegraphics[width=\textwidth]{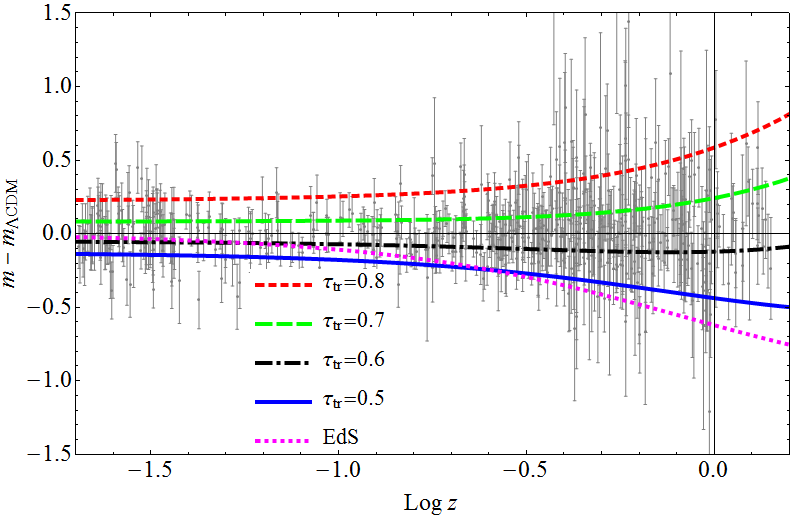}
   \caption{Magnitude residuals versus redshift on a logarithmic scale for the same SNeIa sample (grey points), plotted together with a much more restricted range of plausible models within the theory  (dashed, dash-dotted and solid lines) as well as the Einstein-de~Sitter model (dotted magenta). (The fiducial $\Lambda$CDM model is represented on this plot by the horizontal axis at $m-m_{\text{fid}}=0$).}
\label{fig-magRes}%
\end{figure*}

Figures~\ref{fig-distMod} and \ref{fig-magRes} are plots of distance modulus and magnitude residuals respectively for various values of the transition lookback time $\tau_{\text{tr}}$, as computed from the integral~(\ref{LumDistsigma}) with the limit of integration (\ref{tauzfit}) where $\alpha$ and $h_{\infty}$ are specified by Eqs.~(\ref{nDefn}) and (\ref{hinftyApprox}) respectively.
Comparison with the observational data for actual supernovae in the Union~2.1 catalog (grey points) shows that the value of $\tau_{\text{tr}}$ is tightly constrained.
The magnitude residuals in particular imply that
\begin{equation}
\tau_{tr} = 0.65 \pm 0.05 \; .
\label{SNconstraint}
\end{equation}
The transition to acceleration thus occurred much longer ago than in standard $\Lambda$CDM cosmology, where $\tau_{\text{tr}}=0.34$, as may be easily calculated from $\tau_{\text{tr}}=H_0\int_0^{z_{\text{tr}}}dz/(1+z)H(z)$ where $z_{\text{tr}}=(2\Omega_{\Lambda}/\Omega_M)^{1/3}-1$ \cite{OverduinWesson2008}.
The model of Hong et al. (2008) is not of the standard FRW type, so this observation does not immediately rule it out---but it does suggest a way to proceed.

The universe in the model of Hong et al. \cite{HongEtAl2008} expands at a doubly-exponential rate at early times, so it has no big bang singularity as such.
However, it does become arbitrarily hot and dense as $\tau\rightarrow\infty$, thus retaining the observational successes of the standard big bang theory in principle.
But the time at which this occurs must be sufficiently long ago to accommodate the oldest observed stars.
It should also not have happened many times longer ago than the oldest observed stars; this would be anti-Copernican (implying that we happened to find ourselves in an unusually young corner of the Universe).
We can check these conditions using the range of values for $\tau_{\text{tr}}$ determined by the SNeIa data in Eq.~(\ref{SNconstraint}).

At early times ($\tau\gg1$), the scale factor can be approximated using Eq.~(\ref{scaleFactorFinal}) with Eq.~(\ref{hinftyApprox}) as
\begin{equation}
a(\tau)\approx a_0\exp\left[-\frac{\alpha}{9}\,e^{9(1-\beta e^{-\gamma\tau_{\text{tr}}})(\tau-\tau_{\text{tr}})}\right] \; .
\label{bigBangLimit}
\end{equation}
To invert this expression and obtain an analog for the lookback time to the big bang in this theory, we assume that the universe contains a radiation component (i.e., photons making up the Cosmic Microwave Background or CMB) and that entropy is conserved so that the effective temperature of these photons evolves approximately as $T\propto a^{-1}$, as in standard cosmology \cite{OverduinWesson2008}.
Then Eq.~(\ref{bigBangLimit}) can be re-arranged to give
\begin{equation}
\tau(E)\approx\tau_{\text{tr}}+\frac{\ln\left[9\ln(E/k T_0)/\alpha\right]}{9(1-\beta e^{-\gamma\tau_{\text{tr}}})} \; ,
\label{ageUniverse}
\end{equation}
where $T_0=2.7$~K is the present CMB temperature and $E=kT$ is the energy scale of the universe at temperature $T$.
The age of the universe is then obtained from $\tau_0=\tau(E_{bb})$ where $E_{bb}$ is the energy scale of the big bang.

Putting the allowed values of $\tau_{\text{tr}}$ from the SNeIa constraint~(\ref{SNconstraint}) into Eq.~(\ref{ageUniverse}), we find that $\tau_0=1.5\pm0.1$ for {\em any} value of $E_{bb}$ ranging from 1~MeV (i.e., standard big-bang nucleosynthesis) to $10^{19}$~GeV (Planck energy).
This insensitivity to $E_{bb}$ arises physically from the doubly-exponential nature of the expansion at early times, so that $\tau_0$ goes as the log of the log of $E_{bb}$.
The predicted age of the universe in the theory of Hong et al. \cite{HongEtAl2008} thus lies in the range $t_0 = \tau_0/H_0 = 21 \pm 2$~Gyr, regardless of the details of how radiation is incorporated into the theory.
This is much older than the age of the standard $\Lambda$CDM model (14~Gyr).
More importantly, it is incompatible with the upper limit on the age of the Universe in {\em any\/} cosmological model, based solely on the age of the oldest stars: $\tau_0 = 15 \pm 4$~Gyr \cite{CowanSneden2006}.
The theory is thus effectively ruled out by observation.

It would be of interest to investigate modifications of the theory that might agree better with the data.
The addition of explicit matter or radiation terms in the Lagrangian of the theory would contribute to deceleration and shorten the lifetime of the universe in principle, though this would not likely be a large effect for realistic densities.
Such a modified theory would have additional adjustable parameters, and would likely have to be studied numerically rather than analytically.

The fact that deceleration in this theory is associated with compact extra dimensions is also of interest, suggesting the possibility of a purely ``geometrical'' dark-matter candidate similar to those that have been proposed in the context of Kaluza-Klein gravity \cite{OverduinWesson1997}.
Further study of these questions is left for future work.

\section*{Acknowledgments}

We thank the Towson University Fisher College of Science and Mathematics and the Physics, Astronomy, and Geosciences department for travel support.

\end{document}